\documentclass{article}
\usepackage{graphicx} 
\usepackage{physics}
\usepackage{amsmath,amssymb}
\usepackage{slashed}
\usepackage{bbold}
\usepackage[style=nature]{biblatex} 
\usepackage{authblk}
\usepackage{abstract}

\usepackage{hyperref}
\usepackage{xurl}
\hypersetup{breaklinks=true}

\def\correspondingauthor{\footnote{cameron.cianci@uconn.edu}}

\title{Toward Constructing a Continuous Logical Operator for Error-Corrected Quantum Sensing}
\author[1]{Cameron Cianci \correspondingauthor{}}
\affil[1]{Physics Department, University of Connecticut, Storrs, CT 06269, USA}

\date{}

\begin{document}

\maketitle
\begin{abstract}
Error correction has long been suggested to extend the sensitivity of quantum sensors into the Heisenberg Limit.  However, operations on logical qubits are only performed through universal gate sets consisting of finite-sized gates such as Clifford+T.  Although these logical gate sets allow for universal quantum computation, the finite gate sizes present a problem for quantum sensing, since in sensing protocols, such as the Ramsey measurement protocol, the signal must act continuously. The difficulty in constructing a continuous logical operator comes from the Eastin-Knill theorem, which prevents a continuous signal from being both fault tolerant to local errors and transverse.  Since error correction is needed to approach the Heisenberg Limit in a noisy environment, it is important to explore how to construct fault-tolerant continuous operators.  In this paper, a protocol to design continuous logical z-rotations is proposed and applied to the Steane Code.  The fault tolerance of the designed operator is investigated using the Knill-Laflamme conditions.  The Knill-Laflamme conditions indicate that the diagonal unitary operator constructed cannot be fault tolerant solely due to the possibilities of X errors on the middle qubit.  The approach demonstrated throughout this paper may, however, find success in codes with more qubits such as the Shor code, distance 3 surface code, [15,1,3] code, or codes with a larger distance such as the [11,1,5] code.
\end{abstract}


\section{Introduction}
Quantum sensors have found utility in a variety of fields including commercial applications such as geoscience and mining \cite{2014APS..MARF23004F}.  There have been many recent studies examining the potential utility of error correction to improve the sensitivity of quantum sensors in noisy environments \cite{10.1117/12.2511587, PhysRevLett.112.150802, Shettell_2021, Rojkov_2022, Herrera_Mart__2015, Matsuzaki_2017, Reiter_2017}.
Error correction in quantum sensors promise to surpass the Standard Quantum Limit, where sensitivity scales as $\frac{1}{\sqrt{t}}$, and instead approach the Heisenberg Limit, scaling as $\frac{1}{t}$ \cite{10.1117/12.2511587}. This scaling is the best allowed by the laws of quantum mechanics.

Current studies into quantum error-corrected sensors propose codes which can correct the most prevalent type of noise in a system but are still vulnerable to other local errors \cite{Eastin_2009}.  For an example, \cite{Herrera_Mart__2015} utilized a code to correct relaxation in a quantum magnetometer, but the sensor designed is still vulnerable to single qubit phase errors.  Although the paper proposes mitigating these phase errors by leveraging dynamical decoupling \cite{Viola_1999, Bylander_2011}, the designed sensor will realistically still accumulate uncorrected errors over time from random environmental fluctuations in the magnetic field. Therefore, this design will be reduced to the Standard Quantum Limit on time scales dictated by the strength of this environmental noise \cite{10.1117/12.2511587}.  This can be addressed by using stronger error-correcting codes such as a distance 3 code, which has the ability to correct single qubit errors.

\hfill

A common quantum sensing protocol is the Ramsey measurement protocol described below \cite{Degen_2017}.

\begin{enumerate}
    \item A sensor qubit begins in the state $\ket{0}$.
    \item A Hadamard gate is applied bringing the state to, $H\ket{0} = \ket{+}$.
    \item The signal is applied to the qubit, giving it a signal dependent phase, $P_L(\phi)\ket{+} = P_L(\phi)\times\frac{1}{\sqrt{2}}(\ket{0}+\ket{1}) = \frac{1}{\sqrt{2}}(\ket{0}+e^{i\phi}\ket{1})$.
    \item A Hadamard gate is applied again, bringing the state to $H\frac{1}{\sqrt{2}}(\ket{0}+e^{i\phi}\ket{1}) = \frac{1+e^{i\phi}}{2}\ket{0} + \frac{1-e^{i\phi}}{2}\ket{1}$.
    \item Measuring in the z-basis, the probability of obtaining $\ket{1}$ is $|\frac{1-e^{i\phi}}{2}|^2$, from which $\phi$ can be inferred.
\end{enumerate}

The continuous phase gate $P_L(\phi)$ acts on the computational basis states as,

\begin{equation}
    P_L(\phi) = 
    \begin{bmatrix}
        1 & 0\\
        0 & e^{i\phi}
    \end{bmatrix}
\end{equation}

The Ramsey measurement protocol requires that there is a continuous symmetry around the z-axis of the qubit for $P_L(\phi)$ to be fault tolerantly applied.  As logical gate sets do not typically include any continuous gates, it is not straightforward to apply this protocol directly to an error-corrected logical qubit.  Instead, current error-corrected sensing protocols leave the logical qubit vulnerable to certain local errors, only correcting the most prominent types of error.  For example, these sensors often employ codes such as the bit flip or amplitude damping codes \cite{Herrera_Mart__2015, Matsuzaki_2017, Reiter_2017}. This design choice ultimately allows for transverse operators to generate the signal, for example, magnetic fields in flux tunable superconducting qubits \cite{Herrera_Mart__2015}.

The reason for this difficulty in designing error-corrected quantum sensors fault tolerant to single qubit errors comes from the Eastin-Knill theorem.  This theorem states that no quantum error-correcting code that can correct local errors can also have a continuous symmetry which acts transversely on the qubits \cite{Eastin_2009}.  This is proven by demonstrating that the set of fault tolerant gates on any local error-correcting code is finite and cannot have any continuous symmetries as a continuous symmetry would imply an infinite number of fault tolerant gates.  Since a continuous symmetry is required in many sensing protocols such as Ramsey measurement shown above, the Eastin-Knill theorem complicates the design of error-corrected quantum sensors. This is the reason why current error-corrected quantum sensors leave a degree of freedom uncorrected and therefore preserve a continuous symmetry for the signal.  However, as was proven in \cite{10.1117/12.2511587}, the presence of any noise along this symmetry will make these sensors revert to the Standard Quantum Limit as they will no longer satisfy the HNLS criterion.

The Eastin-Knill theorem uncovers an interesting question in quantum sensing, is it possible to realize continuous logical operators for error-corrected sensing on a logical qubit?  Therefore, our goal in Sections 3 and 4 will be to construct a non-transverse logical phase operator, $P_L(\phi)$, acting on the logical subspace for the purpose of creating quantum error-corrected sensors.  The difficulty constructing this operator is likely what has prevented prior exploration into correcting local errors in quantum sensors.

\section{Arbitrary Diagonal Unitary Gate}
One problem in creating a fault tolerant phase operator is that errors may occur between the gates constructing this operator.  This would increase the number of possible errors, requiring a larger code which can recognize these new error syndromes.  To circumvent this problem, we will consider diagonal unitary operators and demonstrate that they can be built from commuting gates which could be applied simultaneously.  Additionally, restricting the operator to be diagonal greatly reduces its complexity from $(2^{n})^2$ to $2^n$ degrees of freedom.  We will also find the requirements for a creating logical phase gate are simpler when restricted to a diagonal unitary.

This operator can be constructed from a single qubit z-axis rotation gate, $R_Z(\phi)$, controlled by $n$ qubits where $n \in \{0, 1, ... N\}$ with the total number of qubits $N$ (ex. $R_{Z_1}(\phi)$,$C_1R_{Z_2}(\phi)$, $C_1C_3R_{Z_4}(\phi)$ ...).  As these controlled-phase gates are all diagonal and therefore commute, it may be possible to realize them simultaneously and prevent errors from occurring between these gates.  Whether these multi-qubit gates can be created in superconducting quantum circuits is a topic for future investigation, and we will focus only on constructing the logical operators.  

Alternatively, other ways to efficiently create diagonal unitaries have been previously explored \cite{Welch_2014}.  However, these diagonal unitaries are built from non-commuting single and two-qubit gates, increasing the number of distinct errors which can occur.

Next, we will programmatically construct an arbitrary diagonal unitary operator from controlled-phase gates.  We begin by noting that, for any given $N$, there are $2^N$ degrees of freedom in a diagonal unitary, and $2^N-1$ different controlled-phase gates.  We will eliminate the first diagonal entry through the application of a global phase, without any loss of generality.

We can now construct an arbitrary diagonal unitary operator through the following protocol.

\begin{enumerate}
    \item Initialize an array to the Identity, in which we will store the constructed operator, $U_C = \mathbb{1}$.
    \item Let the index $i$ loop through each diagonal entry of the desired operator $U_{d}$.
    \begin{enumerate}
        \item Convert the current index $i$ into binary.  This binary representation shows the basis state on which this entry will act (ex. $i = 9 \implies \ket{1001}$).  
        \item Apply a $R_Z(\phi)$ gate to one of the qubits in a $\ket{1}$ state, which is controlled by all other qubits in a $\ket{1}$ state.  This gate is given a value such that, when applied to the constructed operator, the current diagonal element will obtain the desired phase, 
        $(CC...R_Z(\phi) \times U_C)[i][i] = U_d[i][i]$.  
        
        \item Update the constructed operator with this new gate.  $U_C' = CC...R_Z(a) \times U_C$.
    \end{enumerate}
    \item Return the constructed operator $U_C$, and the phases applied at each index.
\end{enumerate}

This protocol will construct any desired diagonal unitary from commuting controlled-phase gates, as the list of applied phases at each index can be used to determine the gates applied.  The correctness of this protocol can be proven through induction, as each gate affects only the current and later diagonal entries in the constructed operator.  Each diagonal entry has a unique operator which can tune its phase without affecting previously considered entries, except for the first entry which can be removed through an application of a global phase.  This unique operator is found by applying a phase gate controlled by the binary representation of the state corresponding to the index of the entry. Therefore, we can simply construct the desired operator by making greedy decisions at each entry.  Now that the construction of diagonal unitary operators from commuting gates has been stated, we will put forward a simple example to clarify.

\subsection{Constructing a Simple Diagonal Unitary Operator}
Here is an example of this protocol used to construct this unitary $U_d$,

\begin{gather*} 
    \begin{matrix}
        \ket{00} & \ket{01} & \ket{10} & \ket{11}
    \end{matrix}
\end{gather*}
\begin{gather*}
    \begin{pmatrix}
        1 & 0 & 0 & 0\\
        0 & e^{i\phi} & 0 & 0\\
        0 & 0 & e^{i2\phi} & 0\\
        0 & 0 & 0 & 1
    \end{pmatrix}
\end{gather*}

First we loop through each diagonal entry.  For each entry, the value can be changed by the phase gate controlled by the values of $1$ in the binary representation of the index.

\begin{equation}    
U_d \ket{01} = e^{i\phi}\ket{01} \implies R_{Z_1}(\phi)
\end{equation}

which in turn gives our constructed operator (previously initialized to $\mathbb{1}$) as,

\begin{equation}
    U_C = R_{Z_1}(\phi)\times\mathbb{1} =
    \begin{pmatrix}
        1 & 0 & 0 & 0\\
        0 & e^{i\phi} & 0 & 0\\
        0 & 0 & 1 & 0\\
        0 & 0 & 0 & e^{i\phi}
    \end{pmatrix}
\end{equation}

Next we find,

\begin{equation}
    U_d \ket{10} = e^{i2\phi}\ket{10} \implies R_{Z_2}(2\phi)
\end{equation}

This makes our constructed operator,

\begin{equation}
    U_C = R_{Z_1}(\phi)\times R_{Z_2}(2\phi)\times \mathbb{1} =
    \begin{pmatrix}
        1 & 0 & 0 & 0\\
        0 & e^{i\phi} & 0 & 0\\
        0 & 0 & e^{i2\phi} & 0\\
        0 & 0 & 0 & e^{i3\phi}
    \end{pmatrix}
\end{equation}

Lastly, viewing the final entry,

\begin{equation}
    U_d \ket{11} = \ket{11} 
\end{equation}

Currently, our constructed operator gives the value,
\begin{equation}
    U_C \ket{11} = e^{i3\phi} \ket{11}
\end{equation}

This indicates that we must apply $C_1R_{Z_2}(-3\phi)$, giving us the final operator,

\begin{equation}
    U_C = R_{Z_1}(\phi)\times R_{Z_2}(2\phi)\times C_1R_{Z_2}(-3\phi)\times \mathbb{1} = 
    \begin{pmatrix}
        1 & 0 & 0 & 0\\
        0 & e^{i\phi} & 0 & 0\\
        0 & 0 & e^{i2\phi} & 0\\
        0 & 0 & 0 & 1
    \end{pmatrix}
\end{equation}

Which is exactly the desired operator, decomposed into two phase gates and one controlled-phase gate.  As was discussed previously, these gates commute and could therefore be applied simultaneously to prevent errors from occurring between them.

\newpage

\section{Creating a Logical Phase Gate}
Next we want to create a logical phase gate from a diagonal operator.  To start, we must consider the code words of the chosen code on which we will be acting.  Since our operator must function as a logical z-rotation gate, it must satisfy the two conditions,

\begin{equation}
    P_L(\phi) \ket{0}_L = \ket{0}_L
\end{equation}
\begin{equation}
    P_L(\phi) \ket{1}_L = e^{i\phi}\ket{1}_L
\end{equation}

These constraints are straightforward when applied to a diagonal operator, as we simply must ensure that all diagonal elements which are multiplied by the nonzero basis states in $\ket{1}_L$ have a value of $e^{i\phi}$ while all diagonal elements which are multiplied by the nonzero basis states in $\ket{0}_L$ have a value of $1$.

For an example, if the logical eigenstates of a desired code were,

\begin{gather*} 
    \ket{0}_L =\ket{000} + \ket{001} + \ket{010} + \ket{100} =
    \begin{bmatrix}
        1 & 1 & 1 & 0 & 1 & 0 & 0 & 0
    \end{bmatrix}
\end{gather*}
\begin{gather*}
    \ket{1}_L = \ket{111} + \ket{110} + \ket{101} + \ket{011} =
    \begin{bmatrix}
        0 & 0 & 0 & 1 & 0 & 1 & 1 & 1
    \end{bmatrix}    
\end{gather*}

These code words would restrict our diagonal logical phase operator to

\begin{gather*}
    P_L(\phi) =
    \begingroup{\setlength{\tabcolsep}{1.15em}
    \renewcommand*{\arraystretch}{1.5}
        \begin{pmatrix}
            1 & 0 & 0 & 0 & 0 & 0 & 0 & 0\\
            0 & 1 & 0 & 0 & 0 & 0 & 0 & 0\\
            0 & 0 & 1 & 0 & 0 & 0 & 0 & 0\\
            0 & 0 & 0 & e^{i\phi} & 0 & 0 & 0 & 0\\
            0 & 0 & 0 & 0 & 1 & 0 & 0 & 0\\
            0 & 0 & 0 & 0 & 0 & e^{i\phi} & 0 & 0\\
            0 & 0 & 0 & 0 & 0 & 0 & e^{i\phi} & 0\\
            0 & 0 & 0 & 0 & 0 & 0 & 0 & e^{i\phi}
        \end{pmatrix}}
    \endgroup
\end{gather*}

as this diagonal unitary uniquely results in,
\begin{equation}
    P_L(\phi) \ket{0}_L = \ket{0}_L
\end{equation}
\begin{equation}
    P_L(\phi) \ket{1}_L = e^{i\phi}\ket{1}_L
\end{equation}

Through application of the protocol from Section 2, we find this operator can be realized by the simultaneous application of the gates $C_1R_{Z_2}(\phi)$, $C_1R_{Z_3}(\phi)$, $C_2R_{Z_3}(\phi)$, and $C_1C_2R_{Z_3}(-2\phi)$.

\subsection{Ambiguous Entries}
In most codes, however, code words do not include a superposition of every basis state.  This leaves ambiguous or unconstrained degrees of freedom in the constructed operator.  For an example, consider a code with the code words $\ket{0}_L = \ket{00}$ and $\ket{1}_L = \ket{11}$.  This leaves an ambiguous logical phase operator,

\begin{equation}
    P_L(\phi) = 
    \begin{pmatrix}
        1 & 0 & 0 & 0 \\
        0 & a & 0 & 0 \\
        0 & 0 & b & 0 \\
        0 & 0 & 0 & e^{i\phi}
    \end{pmatrix}
\end{equation}

We can therefore tune these variables $a$ and $b$ as needed. For another more applicable example, the Steane code logical states include a superposition of $8$ states out of $128$ basis states.

\begin{multline}
    \ket{0}_L = \frac{1}{\sqrt{8}}(\ket{0000000} + \ket{1010101} + \ket{0110011} + \ket{1100110} \\ + \ket{0001111} + \ket{1011010} + \ket{0111100} + \ket{1101001})
\end{multline}

This logical code word restricts only $8$ of the $128$ diagonal values of the logical operator.  The $\ket{1}_L$ state similarly restricts $8$ more leaving $112$ tunable values.  In the next section, we will consider constraining these values in the Steane code in an attempt to make our logical phase operator fault tolerant through satisfying the Knill-Laflamme conditions.

\section{The Fault Tolerance of Designed Logical Phase Gates}
Now that we can construct a diagonal logical phase operator for any error-correcting code given its logical code words, we may now test if the constructed logical operator is fault tolerant.  This can be done by satisfying the Knill-Laflamme conditions, which are both sufficient and necessary for error correction \cite{girvin2023introduction}.

The Knill-Laflamme conditions for a code with code words $\ket{W_\sigma}$ fault tolerant to errors $K_i = \{K_1 K_2, ... K_n\}$ is,

\begin{equation}
    \bra{W_\sigma } K_l^\dag K_k\ket{W_{\sigma'}} = \alpha_{lk}\delta_{\sigma \sigma'}
\end{equation}

The coefficients $\alpha_{lk}$ must have no dependence on $\sigma$ or $\sigma'$. When considering local errors, the Kraus Operators $K_i$ are the single qubit Pauli gates (X, Y, Z, and I).

Assuming we want to make our logical phase operator fault tolerant, we need to expand the Kraus operators such that we account for errors taking place both before and after the application of the logical operator.  Since we can construct this operator from simultaneously applied commuting gates as shown in Section 2, we will not consider errors occurring between gates constructing the logical operator.  With $P_L(\phi)$ as a logical operator, we need to expand the Knill-Laflamme conditions to the following four equations.

\begin{equation}
    \bra{W_\sigma }P^\dag_L(\phi) K_l^\dag K_k P_L(\phi)\ket{W_{\sigma'}} = \alpha_{lk}\delta_{\sigma \sigma'}
\end{equation}

\begin{equation}
    \bra{W_\sigma }K_l^\dag P^\dag_L(\phi) K_k P_L(\phi)\ket{W_{\sigma'}} = \beta_{lk}\delta_{\sigma \sigma'}
\end{equation}

\begin{equation}
    \bra{W_\sigma }P^\dag_L(\phi) K_l^\dag P_L(\phi)K_k\ket{W_{\sigma'}} = \beta_{lk}\delta_{\sigma \sigma'}
\end{equation}

\begin{equation}
    \bra{W_\sigma }K_l^\dag P^\dag_L(\phi) P_L(\phi) K_k\ket{W_{\sigma'}} = \gamma_{lk}\delta_{\sigma \sigma'}
\end{equation}

A logical phase gate which satisfies these conditions will additionally be fault tolerant to single qubit errors.  The error detection and correction operators can then be derived from the Knill-Laflamme equations \cite{girvin2023introduction}.

We must now return our attention to the tunable elements of the logical phase unitary noted in Section 3.1.  We will attempt to satisfy the Knill-Laflamme conditions shown in equations 16, 17, 18, and 19 by using these values.  

Consider the simple example explored in section 3.1 with code words $\ket{0}_L = \ket{00}$ and $\ket{1}_L = \ket{11}$, in the presence of $X_1$ errors, $K_i = \{X_1\}$.  Equations 16, 17, 18, and 19 require $a = 1$ and $b = e^{i\phi}$, making the following logical phase gate tolerant to these $X_1$ errors,

\begin{equation}
    P_L(\phi) = 
    \begin{pmatrix}
        1 & 0 & 0 & 0 \\
        0 & 1 & 0 & 0 \\
        0 & 0 & e^{i\phi} & 0 \\
        0 & 0 & 0 & e^{i\phi}
    \end{pmatrix}
\end{equation}

As can be seen by applying the protocol from section 2, this is simply a z-axis rotation gate on the second qubit $R_{Z_2}(\phi)$.  Now we will attempt to apply this approach to a more powerful error-correcting code, the Steane Code.

\section{Results in the Steane Code}
The Steane Code is one of the simplest and most well-studied error-correcting codes of distance 3, meaning it can correct any single local error \cite{1996}. Since this code can correct local errors, the Eastin-Knill theorem requires the signal to be non-transverse.  However, the approach designed in Sections 2, 3, and 4 is not restricted to transverse operators, and therefore is not forbidden from creating a fault tolerant continuous operator.  We will now apply our process for making logical phase gates to the Steane Code. 

Using the code words of the Steane code (shown below), we restrict the corresponding diagonal elements of our operator as shown in Section 3.
\begin{multline}
    \ket{0}_L = \frac{1}{\sqrt{8}}(\ket{0000000} + \ket{1010101} + \ket{0110011} + \ket{1100110} \\ + \ket{0001111} + \ket{1011010} + \ket{0111100} + \ket{1101001})
\end{multline}
\begin{multline}
    \ket{1}_L = \frac{1}{\sqrt{8}}(\ket{1111111} + \ket{0101010} + \ket{1001100} + \ket{0011001} \\ + \ket{1110000} + \ket{0100101} + \ket{1000011} + \ket{0010110})
\end{multline}

The diagonal elements of the desired logical operator $U_d$ must behave as,

\begin{equation}
    U_d \ket{0}_L = \ket{0}_L
\end{equation}
\begin{equation}
    U_d \ket{1}_L = e^{i\phi}\ket{1}_L
\end{equation}

This leaves $112$ unconstrained degrees of freedom in the logical phase operator. However, when applying the Knill-Laflamme conditions as shown in Section 4, it is found that the following condition is unsatisfiable.

\begin{equation}
     \bra{W_0}X_3P^\dag_L(\phi)X_3P_L(\phi)\ket{W_0} = \bra{W_1}X_3P^\dag_L(\phi)X_3P_L(\phi)\ket{W_1}
\end{equation}

This can be interpreted as an error on the middle qubit of the Steane code before $P_L(\phi)$, which is indistinguishable from an error after the application of $P_L(\phi)$ when restricting $P_L(\phi)$ to be diagonal. 
 
More precisely, the value of $\beta$ in $\bra{W_\sigma }X_3 P^\dag_L(\phi) X_3 P_L(\phi)\ket{W_{\sigma'}} = \beta\delta_{\sigma \sigma'}$ changes sign based on the value of $\sigma$ and $\sigma'$.  This indicates that an $X_3$ error before $P_L(\phi)$ changes the state differently than an $X_3$ error after $P_L(\phi)$, but both errors are recognized by the same error syndromes and are indistinguishable.  Therefore, this approach unfortunately does not succeed in creating a logical phase operator in the Steane code.

\section{Future Directions}
Although this approach does not succeed at creating a fault tolerant logical phase operator in the Steane code, it is possible that this approach may succeed when applied to larger codes.  

The $[11,1,5]$ code is typically tolerant to two qubit errors and may still be tolerant to a single local error when designing a continuous logical operator.  Due to the increased code distance, it is possible that even if an error propagates through the multi-qubit gates of the logical operator, the larger distance of this code may still be able to correct these errors.

Additionally, the Shor Code and the distance 3 surface code \cite{Fowler_2012, Krinner_2022} have a higher qubit count, which may be able to accommodate more error syndromes and correct more errors than the Steane code.  Therefore, diagonal operators in these codes may still be fault tolerant. 

Furthermore, the diagonal restriction on the logical operator could be lifted to allow for more tunable degrees of freedom.  However, when considering this, it is important to remember that the gates composing this logical operator may not commute, meaning gates may not be applied simultaneously and errors may occur between them.  This will likely result in further issues for the fault tolerance of the logical operator.


Also, as the Solovay-Kitaev theorem allows for universal computation to be achieved from Clifford+T gates in $O(m\log^c(m/\epsilon))$ \cite{nielsen00}, a logarithmic speedup may be possible using a continuous operator instead of the T gate.  However, a logarithmic speedup is not often significant compared to the quadratic or exponential speedups commonly provided by quantum algorithms.

One of the most interesting outcomes may be that it is impossible to construct a fault tolerant continuous logical operator in error-correcting codes that can be decomposed into physically realizable gates.  If this is the case, it would potentially indicate the presence of new and interesting theorems for error correction and quantum sensing.

\printbibliography

\end{document}